# Observation of double hysteresis in CoFe$_2$O$_4$/MnFe$_2$O$_4$ core/shell nanoparticles and its contribution to AC heat induction

Jie Wang,[a] Hyungsub Kim,[a] HyeongJoo Seo,[b] Satoshi Ota,[c] Chun-Yeol You,[b] Yasushi Takemura,[d] and Seongtae Bae[a*]

[a]Nanobiomagnetics and Bioelectronics Laboratory (NB$^2$L), Department of Electrical Engineering, University of South Carolina, Columbia, SC, 29208, USA
[b]Department of Emerging Materials Science, Daegu Gyeongbuk Institute of Science & Technology, Daegu, 42988, Republic of Korea
[c]Department of Electrical and Electronic Engineering, Shizuoka University, Hamamatsu 432-8561, Japan
[d]Department of Electrical and Computer Engineering, Yokohama National University, Yokohama, 240-8501, Japan

*Corresponding author:
Email: bae4@cec.sc.edu

**Abstract**

Magnetic core/shell nanoparticles are promising candidates for magnetic hyperthermia due to its high AC magnetic heat induction (specific loss power (SLP)). It's widely accepted that magnetic exchange-coupling between core and shell plays the crucial role in enhancing SLP of magnetic core/shell nanoparticles. However, the physical contribution of exchange coupling to SLP has not been systematically investigated, and the underlying mechanism remains unclear. In this study, magnetic hard/soft $CoFe_2O_4/MnFe_2O_4$ and inverted soft/hard $MnFe_2O_4/CoFe_2O_4$ core/shell nanoparticles were synthesized, systematically varying the number of shell layers, to investigate the physical contribution of internal bias coupling at the core/shell interface to AC heat induction (SLP). Our results show that a unique magnetic property, double-hysteresis loop, was present and clearly observed, which was never reported in previous core/shell research literature. According to the experimentally and theoretically analyzed results, the double-hysteresis behavior in core/shell nanoparticles was caused by the difference of magnetic anisotropy between core and shell materials, separated by a non-magnetic interface. The enhanced SLP and and maximum temperature rise ($T_{AC,max}$) of core/shell nanoparticles are attributed to the optimized magnetic anositropy, AC magnetic softness and double hysteresis behavior due to the internal bias coupling. These results demonstrate that the rational design capabilities to separately control the magnetic anisotropy, AC/DC magnetic properties by varying the volume ration between core and shell and by switching hard or soft phase materials between core and shell are effective modalities to enhance the AC heat induction of core/shell nanoparticles for magnetic nanoparticle hyperthermia.

## 1. Introduction

Magnetic nanoparticle hyperthermia (MNH) using iron oxide nanoparticles (IONPs) has attracted significant attention as a potential modality for cancer treatment due to the low side effect and high treatment efficacy [1-3]. However, several challenges, especially low AC magnetic heat induction or specific loss power (SLP) generated by IONPs under the biologically safe and physiologically tolerable range of AC magnetic field ($H_{AC}$) ($H_{AC,\,safe} < 1.8 \times 10^9$ A m$^{-1}$ s$^{-1}$, $f_{appl} <$ 120 kHz, $H_{appl} < 190$ Oe (15.2 kA m$^{-1}$)), limit the progress of MNH in cancer clinical translation [4-6]. Accordingly, extensive research activities have been conducted over the past decades to optimize the SLP of IONPs, which is controlled by many factors, including intrinsic (magnetic moment, magnetic anisotropy, etc.) and extrinsic (particle size, shape, etc.) properties of IONPs as well as the frequency and intensity of the externally applied AC magnetic field, to generate higher AC heat induction at the $H_{AC,\,safe}$ for cancer therapeutics. These efforts include: (1) optimizing the chemical synthesis routes to tailor the composition, size, shape, and size distribution of IONPs, to control the magnetocrystalline anisotropy, surface anisotropy, shape anisotropy, magnetic dipolar interaction, respectively [7-10], (2) artificially employing doping techniques (shallow doping, dual-doping, etc.) using various magnetic and/or non-magnetic cations to enhance the AC/DC magnetic properties of IONPs [11, 12], (3) constructing new types of physical structures, e.g., core/shell structure, magnetic vortex structure, etc., to modify the magnetic anisotropy of IONPs [13-15].

Among these efforts, magnetic core/shell nanostructures are promising candidates enabling to generate high SLP for magnetic hyperthermia due to its enhanced magnetic moment and tunable magnetic anisotropy. The magnetic core/shell structure is a particular class of MNPs consisting of a ferromagnetic (or antiferromagnetic) core and an antiferromagnetic (or ferromagnetic) shell, such as Co/CoO, Ni/NiO etc., or a magnetically soft (hard) core and a magnetically hard (soft) shell, e.g., $CoFe_2O_4/MnFe_2O_4$, $FePt/Fe_3O_4$, etc [13, 16-19]. Such complex structures have become of great interest as they provide an additional degree of freedom and open a new route to design novel magnetic nanomaterials for MNH application due to the combination of distinct properties of the diverse constituents in a single component. More importantly, the advantages of the core/shell structure lie not only in multifunctionality, but also in the possibility to improve and tune the single-phase properties using the interactions between the different components [20-22]. Magnetically hard/soft $CoFe_2O_4/MnFe_2O_4$ core/shell nanoparticle is a classical and appealing architecture, which shows high AC magnetic heat induction power (SLP), and also exhibits superior biocompatibility and therapeutic effect for cancer treatment in MNH [9, 13]. Most research studies have attributed the high SLP of magnetic core/shell nanoparticles, including $CoFe_2O_4/MnFe_2O_4$, to the magnetic exchange coupling between the core and shell [13, 23-25]. However, the physical contribution of magnetic exchange coupling to SLP for magnetic hyperthermia was not systematically investigated, and the origin of exchange coupling for the high SLP remains poorly understood. More importantly, little attention has been given to cation diffusion across the interface between core and shell phases, which can be observed in magnetic thin film systems or semiconductor materials due to the differences in diffusion rate and solid solubility in two different phases [26, 27].

In this study, magnetic hard/soft $CoFe_2O_4/MnFe_2O_4$ core/shell IONPs were synthesized with systematically tunning the number of shell layers to investigate the physical contribution of internal bias coupling at the interface to the AC heat induction (SLP). To clarify the distinct magnetic properties of the core/shell IONPs, a comparative study using the inverted magnetic soft/hard $MnFe_2O_4/CoFe_2O_4$ core/shell IONPs was conducted. Structural confirmation of the core/shell structure was achieved using scanning transmission electron microscope coupled with energy dispersive x-ray spectroscopy (STEM-EDX) to map the chemical composition at the nanoscale. Notably, a unique magnetic property, double-hysteresis loop, was present and observed, which was never reported in previous core/shell research literature. Micromagnetic simulation was used to provide the physical insight of the double-hysteresis behavior in core/shell IONPs. The relationship between double-hysteresis behavior and AC magnetic heat induction of core/shell IONPs was fully discussed.

## 2. Materials and Methods

### 2.1 Synthesis of $CoFe_2O_4$/$MnFe_2O_4$ and $MnFe_2O_4$/$CoFe_2O_4$ Core/Shell IONPs with Varying Shell Numbers

$CoFe_2O_4$/$MnFe_2O_4$ core/shell IONPs were synthesized by a modified thermal decomposition method combined with a seed-mediate growth technique [18, 28]. Iron acetylacetonate $Fe(acac)_3$, cobalt acetylacetonate ($Co(acac)_2$) and manganese acetylacetonate ($Mn(acac)_2$) act as chemical precursors. The core $CoFe_2O_4$ and $MnFe_2O_4$ nanoparticles were synthesized using the previous experimental protocol. The synthesis strategies for constructing core/shell IONPs are illustrated in Figure 1. To prepare 6.5 nm $CoFe_2O_4$ core coated with 1 nm $MnFe_2O_4$ shell, 80 mg of $CoFe_2O_4$ nanoparticles were mixed and magnetically stirred with $Mn(acac)_2$ (2 mmol), 1,2-hexadecanediol (10 mmol), oleic acid (6 mmol), oleylamine (6 mmol) and benzyl ether (20 mL) in 250 mL round-bottom flask as described in Figure 1. The mixed solution was heated up to 100°C in 15 min and held for 30 min. Then, the mixture was heated to 200°C for 25 min (~ 8 °C/min) and maintained for 2 h under a flow of nitrogen at the flow rate of 100 mL/min. Subsequently, the solution was further heated to 300 °C for 25 min (~ 4 °C/min) and maintained for 1 h. After removing the heat source, the reaction mixture was cooled down to room temperature. The black-colored mixture was precipitated and separated via centrifugation to remove any undispersed residue. For the synthesis of $CoFe_2O_4/MnFe_2O_4$ core/shell IONPs with 2, 3 and 4 shells, the same fabrication process was performed using the previously synthesized nanoparticles as seeds. The same fabrication procedure was used to synthesize the inverted $MnFe_2O_4/CoFe_2O_4$ core/shell IONPs.

**2.2 Structural and morphological characterization of core/shell IONPs**

The crystal structure of $CoFe_2O_4/MnFe_2O_4$ and $MnFe_2O_4/CoFe_2O_4$ core/shell IONPs was investigated using a Cu-K$_\alpha$-radiated X-ray diffractometer (XRD, Miniflex, Rigaku, Japan). The particle shape/size of the synthesized core/shell IONPs was characterized using a transmission electron microscope (TEM, HT7800, Hitachi, Japan). The size and size distribution of each sample were obtained by measuring the average diameter of more than 300 particles using images collected from different parts of grid. High-resolution TEM (HR-TEM) images and chemical composition of core/shell IONPs were carried out in scanning transmission electron microscopy (STEM) mode. STEM and energy-dispersive X-ray spectroscopy (EDX) were performed on an aberration-corrected JEOL JEM-ARM200CF operated at 200 kV, which can achieve a spatial resolution of ~73 pm. The ARM200CF is equipped with a cold-field emission gun, providing an energy resolution of 350 meV. EDX spectra were collected using an Oxford X-Max 100TLE windowless silicon drift EDX detector.

**2.3 AC magnetic heat induction characteristics of core/shell IONPs**

The AC heat induction characteristics of $CoFe_2O_4/MnFe_2O_4$ and $MnFe_2O_4/CoFe_2O_4$ core/shell IONPs were analyzed using an AC magnetic field induction system consisting of AC coils, capacitors, DC power supplies, and wave generators. The system was designed and developed to apply the $H_{appl}$, and $f_{appl}$ changed from 80 to 160 Oe, and 50 to 360 kHz, respectively.

The sample tubes containing the core/shell IONPs were placed in the center of AC magnetic coil to characterize the AC magnetic heat induction properties. The externally applied AC magnetic field was fixed at the biologically safe range of $H_{appl}$ =140 Oe and $f_{appl}$ = 100 kHz for this study. The AC heat induction temperature of the core/shell IONPs during measurement was detected using a fiber-optic thermometer. The measurement for each sample was conducted 3 times at the same conditions. The SLP values of the IONPs were calculated using the measured AC heat induction curves based on SLP (W/g) = $\frac{C}{m} \cdot \frac{\Delta T}{\Delta t}$, where $C$ is specific heat capacity, $m$ is mass of magnetic nanoparticles and $\Delta T/\Delta t$ is the initial slope of the curve of temperature versus time at the first 30s of the heating curves.

## 2.4 AC and DC magnetic properties of core/shell IONPs

The DC magnetic properties, including magnetic major and minor hysteresis loops of $CoFe_2O_4/MnFe_2O_4$ and $MnFe_2O_4/CoFe_2O_4$ core/shell IONPs, were measured using a vibrating sample magnetometer (VSM, 7400S, Lake Shore, USA). The applied DC magnetic fields were swept at ±5000 Oe for major loop and ±150 Oe for minor loop, respectively. The AC magnetic properties of the core/shell IONPs were analyzed using an AC susceptometer to evaluate AC magnetic softness and out-of-phase susceptibility ($\chi''$) at the fixed $H_{appl}$ = 140 Oe and $f_{appl}$ =100 kHz, which is the same condition as the AC heat induction measurement. The blocking temperature ($T_B$) of the core/shell IONPs was determined using zero-field-cooling/field-cooling (ZFC/FC) curves at the temperature changed from 4.2 K to 304 K, to evaluate the magnetic anisotropy and superparamagnetic status of the core/shell IONPs.

## 3. Results and discussion

## 3.1 Morphology, structure and chemical composition of $CoFe_2O_4/(MnFe_2O_4)_n$ and $MnFe_2O_4/(CoFe_2O_4)_n$ core/shell IONPs

Monodispersed magnetic core/shell IONPs, consisting of a $CoFe_2O_4$ core with several $MnFe_2O_4$ shells $CoFe_2O_4/(MnFe_2O_4)_n$ and a $MnFe_2O_4$ core with several $CoFe_2O_4$ shells $MnFe_2O_4/(CoFe_2O_4)_n$, were synthesized using a combination of thermal decomposition and seed-mediated growth. TEM images, shown in Figure 2(a), 2(b) and Figure S2, revealed that all core/shell IONPs exhibited a well-separated spherical shape and narrow size distribution with a standard deviation $\sigma < 10\%$ after fitting the size histogram. The core/shell IONPs had a core diameter of 6.5 nm, and the number of shell layers was systematically controlled from 1 to 4 with each shell layer determined to be approximately 1 nm in thickness, resulting in the shell thickness ranging from 1 to 4 nm. A systematical study by high-resolution microscopy (HR-TEM and HR-STEM) has been conducted on the synthesized core/shell IONPs. Figure 2(c)-(f) show the HR-STEM images of $CoFe_2O_4/(MnFe_2O_4)_3$ and $MnFe_2O_4/(CoFe_2O_4)_3$ (with 3 shell layers). As shown in Figure 2(c) and (d), the core/shell architecture of $CoFe_2O_4/(MnFe_2O_4)_3$ and $MnFe_2O_4/(CoFe_2O_4)_3$ IONPs can be visualized by determining the different phase contrast and discontinuous atomic lattice fringes. The interlayer distance shown in Figure 2(e) and (f) confirmed the presence of the cubic spinel structure in agreement with the XRD data (Figure S1).

The high-angle annular dark field scanning transmission electron microscopy (HAADF-STEM) and energy dispersive X-ray spectroscopy (EDS) mapping micrographs shown in Figure 3(a)-(f) revealed the homogenous distribution of Mn atoms on the surface of the $CoFe_2O_4$ core. Similarly, Figure 4(a)-(f) show the distribution of Co atoms on the surface of the $MnFe_2O_4$ core. To verify the formation of core/shell architecture, the line-scanned STEM-EDS was used to analyze the chemical composition at the core and shell sections of $CoFe_2O_4/(MnFe_2O_4)_n$ and $MnFe_2O_4/(CoFe_2O_4)_n$ core/shell IONPs shown in Figure 3(f), 4(f) and Figure S3-S8. As shown in $CoFe_2O_4/MnFe_2O_4$ and $MnFe_2O_4/CoFe_2O_4$ core/shell IONPs with 1 shell (Figure S3 and S6), the element (Co or Mn) was scatteringly distributed around the core, and some facets of the $CoFe_2O_4$ and $MnFe_2O_4$ core were not fully covered by the shell of $MnFe_2O_4$ and $CoFe_2O_4$, respectively, due to the low thickness and poor crystalline of the shell which has been verified by previous reports [23, 29]. In contrast, as additional shells were sequentially grown through multi-step chemical reaction, the core was gradually fully covered by the shell material. This leads to the formation of a well-defined core/shell interface and enhanced crystallinity, as evidenced by the sharper contrast in Figure 2(c)-(f).

## 3.2 DC magnetic properties of $CoFe_2O_4/(MnFe_2O_4)_n$ and $MnFe_2O_4/(CoFe_2O_4)_n$ core/shell IONPs

The major, and minor DC M-H loops of $CoFe_2O_4/(MnFe_2O_4)_n$ and $MnFe_2O_4/(CoFe_2O_4)_n$ core/shell IONPs were measured at room temperature using VSM at the sweeping field of ±5000 Oe, and ±150 Oe, respectively, to investigate the effect of the various shell layers on the changes of DC magnetic properties. The DC magnetic properties of the core/shell IONPs including saturation magnetization ($M_s$), coercivity ($H_c$), and initial magnetic susceptibility ($\chi_0$) are summarized in Table 1. As shown in Figure 5 and Table 1, $CoFe_2O_4/MnFe_2O_4$ and $MnFe_2O_4/CoFe_2O_4$ core/shell IONPs exhibited the enhanced $M_s$ compared to the single core component $CoFe_2O_4$ and $MnFe_2O_4$, which can be attributed to the exchange coupling in the interface between core and shell. Moreover, the $M_s$ value keeps increasing with synthesizing more $MnFe_2O_4$ shell layers on the surface of $CoFe_2O_4$. The reason could be that $MnFe_2O_4$ has a low magnetocrystalline anisotropy (high magnetic softness) because of weak electron spin-orbit coupling of the Mn cations [18, 30]. In contrast, the $M_s$ gradually decreases with forming more $CoFe_2O_4$ shell layers on the surface of $MnFe_2O_4$ IONPs due to the magnetic semi-hardness of $CoFe_2O_4$. More interesting behavior of double-hysteresis loops of $CoFe_2O_4/(MnFe_2O_4)_n$ and $MnFe_2O_4/(CoFe_2O_4)_n$ core/shell IONPs (n=1, 2, 3) was observed in DC minor hysteresis loops (Figure 5, bottom). Different from previously reported results when two different magnetic IONPs are physically mixed together, a kink behavior can be observed in hysteresis loops due to the lack of exchange interactions between the two materials [31-33]. The double-hysteresis behavior was resulted from the difference of magnetic anisotropy between core and shell materials, separated by a non-magnetic material [34, 35]. This is the first time of this particular feature was observed in $CoFe_2O_4/(MnFe_2O_4)_n$ and $MnFe_2O_4/(CoFe_2O_4)_n$ core/shell IONPs. In $CoFe_2O_4/MnFe_2O_4$ core/shell IONPs, the double hysteresis loops were very weak to be observed in the case of one $MnFe_2O_4$ shell layer due to the small thickness of shell layer and poor crystallinity, which do not provide sufficient magnetic anisotropy. When two or three shell layers were synthesized onto the surface of core, the double hysteresis loops become more prominent, as the magnetic anisotropy of the shell materials began to compete with that of the core $CoFe_2O_4$. Moreover, with an increasing number of shell layers, the magnetic anisotropy of the shell materials ($MnFe_2O_4$) dominated the core/shell system, as evidenced by the large thickness and high-volume fraction of the shell layers. Similarly, the double hysteresis loops can be easily observed in the early stages of synthesizing two $CoFe_2O_4$ shells onto the surface of core $MnFe_2O_4$. That is because $CoFe_2O_4$ has a higher magnetocrystalline anisotropy than $MnFe_2O_4$, and it can exhibit efficient magnetic anisotropy energy ($K_uV$) at small size to compete with the magnetization of core $MnFe_2O_4$ under externally applied DC magnetic field. With synthesizing more shell layers on the surface of core, both of $CoFe_2O_4/(MnFe_2O_4)_n$ and $MnFe_2O_4/(CoFe_2O_4)_n$ core/shell IONPs tend to show ferrimagnetic or ferromagnetic behavior due to the increase of particle size. These results can be also verified by measuring ZFC-FC curves shown in Figure 6.

This unique double hysteresis behavior observed in this study was attributed to the difference of magnetic anisotropy between core and shell materials, separated by a non-magnetic material. The interface physics is an interesting but great challenging topic in the research of nanoparticles, especially in core/shell IONPs. Some of the previous reports have provide direct and indirect evidence to verify the inter-mixing interface, graded magnetic interface, and paramagnetic/nonmagnetic interlayer in the core/shell IONPs [36-38]. In this study, micromagnetic simulation using the OOMMF software package [39] as an indirect method was employed to investigate the magnetic configuration of core/shell IONPs in the presence and absence of a non-magnetic interlayer between core and shell. As shown in Figure S9 and S11, the double-hysteresis loops can be clearly observed in $CoFe_2O_4$/nonmag/$MnFe_2O_4$ and $MnFe_2O_4$/nonmag/$CoFe_2O_4$, while it is very weak in the $CoFe_2O_4$/$MnFe_2O_4$ and $CoFe_2O_4$/$MnFe_2O_4$ core/shell IONPs without nonmagnetic interlayer (Figure S10 and S12). Furthermore, the overall magnetic behavior of core/shell IONPs tends to be dominated by shell materials due to the high-volume fraction. In addition, the nonmagnetic interlayer between core and shell may be a form of metallic alloy or mattallic compound caused by the cation diffusion during the multi-step synthesis process. The combination of DC magnetic properties measurement, STEM-EDS, and micromagnetic simulation indirectly speculate that the nonmagnetic interlayer caused by cation diffusion may be a means of controlling the double-hysteresis behavior in core/shell IONPs, and it certainly deserves further investigation to provide direct physical evidence.

Figure 6 shows the temperature dependence of the ZFC and FC magnetization of the core/shell IONPs. The blocking temperature ($T_B$) of both $CoFe_2O_4$/($MnFe_2O_4$)$_n$ and $MnFe_2O_4$/($CoFe_2O_4$)$_n$ core/shell IONPs shifts to higher temperature with synthesizing more shell layers. However, the $T_B$ value rises much faster for $MnFe_2O_4/(CoFe_2O_4)_n$ than that for $CoFe_2O_4/(MnFe_2O_4)_n$ core/shell IONPs. Such a discrepancy can be understood by the Stoner-Wohlfarth model [30], where the magnetic anisotropy energy ($E_A=K_u \cdot V$) is determined by the magnetocrystalline anisotropy constant (or uniaxial anisotropy, $K_u$) and volume (V). Since $T_B$ is correlated to $E_A$, it depends on $K_u$ and V accordingly. $T_B$ increases in general as the total volume increases with adding more shell volume. The increase of $T_B$ is more significant in $MnFe_2O_4/(CoFe_2O_4)_n$ core/shell IONPs due to the enhanced magnetocrystalline anisotropy from the addition of $CoFe_2O_4$. This is because $K_u$ ($9.08 \times 10^4$ J/m$^3$) of $CoFe_2O_4$ is close to several orders of magnitude larger than $K_u$ ($1.32 \times 10^4$ J/m$^3$) of $MnFe_2O_4$. The change is very profound even if the shell thickness is only ~ 1 nm (1 shell layer). Moreover, $MnFe_2O_4/(CoFe_2O_4)_n$ IONPs more easily exhibit ferromagnetic properties than $CoFe_2O_4/(MnFe_2O_4)_n$ core/shell IONPs due to the large magnetocrystalline anisotropy of the $CoFe_2O_4$ shell.

## 3.3 AC magnetic heat induction characteristics of $CoFe_2O_4/(MnFe_2O_4)_n$ and $MnFe_2O_4/(CoFe_2O_4)_n$ core/shell IONPs

The AC heat induction characteristics of $CoFe_2O_4/(MnFe_2O_4)_n$ and $MnFe_2O_4/(CoFe_2O_4)_n$ core/shell IONPs were measured at the $H_{ac,safe} = 1.12 \times 10^9$ A m$^{-1}$ s$^{-1}$ ($f_{appl}$=100 kHz, $H_{appl}$=140 Oe (11.2 kA m$^{-1}$)). The AC heating curves and the corresponding $T_{AC, max}$ as well as SLP of $CoFe_2O_4/(MnFe_2O_4)_n$ and $MnFe_2O_4/(CoFe_2O_4)_n$ core/shell IONPs are shown in Figure 7(a)-(c), and Figure 7(i)-(iii), respectively. The SLP was calculated using the measured AC heat induction curves of core/shell IONPs dispersed in hexane. As expected, core/shell IONPs show the higher $T_{AC, max}$ and SLP than the single phase $CoFe_2O_4$ and $MnFe_2O_4$ IONPs. The similar results were found by previous reports and the physical mechanism was explained by the enhanced magnetic moment μ and exchange interaction in core/shell IONPs [7, 13, 14]. Both of $CoFe_2O_4/(MnFe_2O_4)_3$ and $MnFe_2O_4/(CoFe_2O_4)_3$ core/shell IONPs exhibited highest $T_{AC,max}$ and SLP, and then decreased by synthesizing the forth shell layer. It can be understood that the enhanced SLP of $CoFe_2O_4/(MnFe_2O_4)_n$ IONPs was attributed to the increased magnetic moment μ by synthesizing magnetic soft $MnFe_2O_4$. However, the SLP of $MnFe_2O_4/(CoFe_2O_4)_n$ IONPs shows an inverse relation with the magnetic moment shown in Figure 7(iii) and Table 1. This is apparently in contrast with the results reported in the literature for similar systems, and such a discrepancy could be related to AC magnetic softness and size-dependent hysteresis loos. As our previous report [7], DC magnetic softness including magnetic moment (μ) and initial susceptibility ($\chi_0$) partially contributes to the enhancement of SLP, while magnetic anisotropy ($K_u$)-dependent AC magnetic softness, especially out-of-phase susceptibility ($\chi''$) is the most dominant parameter to affect $T_{AC,max}$ (SLP), as described by Eq. (1).

$$P_{relaxation\ loss} = \pi\mu_0 H^2_{AC,appl}(\omega t)f_{appl} \cdot \chi_0 \frac{2\pi f_{appl}\tau_{eff}}{1+(2\pi f_{appl}\tau_{eff})^2} \qquad (1)$$

where $P_{relaxation\ loss}$ is the AC magnetic heat induction power, $\tau_N$ is Neel relaxation time constant, $\tau_B$ is Brownian relaxation time constant, which can be neglected in solid-state IONPs, $\tau_0$ is spin relaxation time constant ($\approx 10^{-9}$ sec.), $\chi'$ is in-phase magnetic susceptibility (not $f_{appl}$-dependent), $\delta_m$ is loss angle, and $\mu=1+4\pi\chi$ is magnetic permeability. The SLP in core/shell IONPs is usually explained in terms of size (or particle volume V), saturation magnetization ($M_s$) and magnetic anisotropic ($K_u$) caused by exchange interaction. Nonetheless, AC magnetic softness ($\chi''$) is more dominant for AC heat induction of IONPs.

### 3.4 AC magnetic properties of $CoFe_2O_4/(MnFe_2O_4)_n$ and $MnFe_2O_4/(CoFe_2O_4)_n$ core/shell IONPs

In order to investigate the effect of various shell layers on the AC magnetic softness ($\chi''$) and AC heat induction of core/shell IONPs, the AC magnetic properties of $CoFe_2O_4/(MnFe_2O_4)_n$ and MnFe$_2$O$_4$/(CoFe$_2$O$_4$)$_n$ core/shell IONPs were investigated using an AC susceptometer at the fixed applied frequency $f_{appl}$ = 100 kHz. As shown in Figure 8, with synthesizing more MnFe$_2$O$_4$ shells onto the CoFe$_2$O$_4$ core, the hysteresis area $A_{AC,hysteresis}$ of CoFe$_2$O$_4$/(MnFe$_2$O$_4$)$_n$ core/shell IONPs was increased up to n=3 (the largest $A_{AC,hysteresis}$), and then decreased by synthesizing the 4th shells. As previous report described, a large $A_{AC,hysteresis}$ indicates that it has a high $\chi''$ and corresponding the large SLP, which is coincident with the previous reports [7, 11, 12]. In contrast, the AC hysteresis loop area $A_{AC,hysteresis}$ of MnFe$_2$O$_4$/(CoFe$_2$O$_4$)$_n$ core/shell IONPs was decreased by synthesizing more CoFe$_2$O$_4$ shells onto the MnFe$_2$O$_4$ core, which indicates MnFe$_2$O$_4$/(CoFe$_2$O$_4$)$_n$ core/shell IONPs exhibit low AC magnetic softness. However, the AC magnetic heat induction including $T_{AC,max}$ and SLP of MnFe$_2$O$_4$/(CoFe$_2$O$_4$)$_n$ core/shell IONPs show the inverse relation with $\chi''$ (AC magnetic softness), which was caused by the contribution of hysteresis loss of MnFe$_2$O$_4$/(CoFe$_2$O$_4$)$_n$ core/shell IONPs to the AC heat induction shown in Figure 5.

## 4. Conclusions

A combination of hard phase CoFe$_2$O$_4$ and soft phase MnFe$_2$O$_4$ as the CoFe$_2$O$_4$/(MnFe$_2$O$_4$)$_n$ and MnFe$_2$O$_4$/(CoFe$_2$O$_4$)$_n$ core/shell IONPs have been synthesized via thermal decomposition and seed-mediated method, and the shell layers have been tuned from n=1 to n=4 by the multi-step chemical reaction. and their magnetic properties as well as AC magnetic heat induction have been systematically investigated. Both HR-TEM and STEM-EDS results have confirmed the formation of the core/shell architecture. An unique magnetic characteristic, double-hysteresis loop, was clearly observed. According to the experimentally and theoretically analyzed results, the double-hysteresis loop behavior was caused by the difference of magnetic anisotropy between core and shell materials, and separated by a nonmagnetic material due to the cation diffusion in the interface. Both of the $CoFe_2O_4/(MnFe_2O_4)_n$ and $MnFe_2O_4/(CoFe_2O_4)_n$ core/shell IONPs shown enhanced SLP ($T_{AC,max}$) compared to the single phase core $CoFe_2O_4$ or $MnFe_2O_4$ due to the internal bias coupling. In addition, the shell layer was found to effectively affect the AC/DC magnetic properties and AC heat induction characteristics. The enhanced SLP and $T_{AC,max}$ are resulted from the optimization of magnetic anositropy, AC magnetic softness and double hysteresis behavior due to the internal bias coupling. These results evidently demonstrate that the rational design capabilities to separately control the magnetic anisotropy, AC/DC magnetic properties by varying the volume ration between core and shell and by switching hard or soft phase materials between core and shell are effective modalities to enhance the AC heat induction of core/shell IONPs for magnetic nanoparticle hyperthermia.

**Acknowledgements**

The research was supported by Neo-Nanomedics Co. Ltd., in South Korea (Grand No. 1550LA04) and was partially supported by SPARC Graduate Research Grant from the Office of the Vice President for Research at the University of South Carolina. In addition, National Research Foundation of Korea (Grand No. 2018R1A5A1025511) from Daegu Gyeongbuk Institute of Science & Technology (DGIST) supported the ZFC-FC measurement.

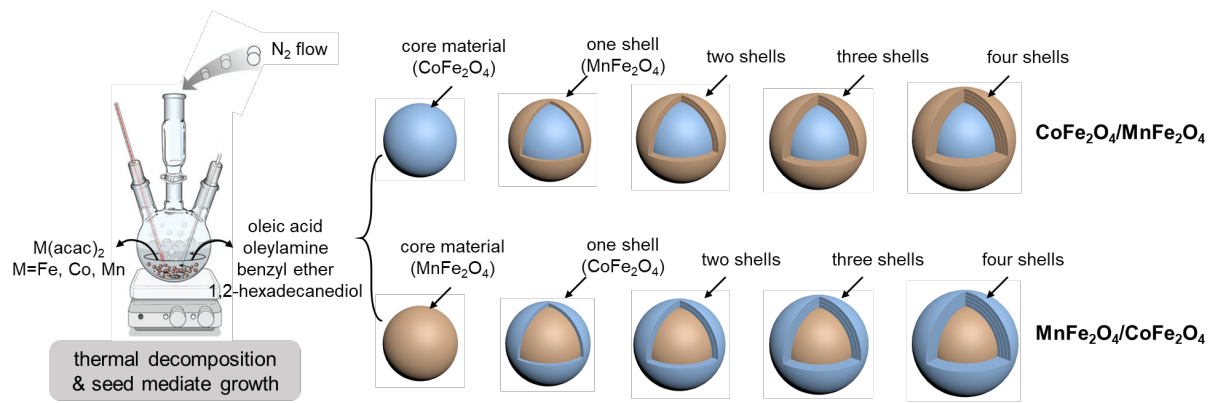

Figure 1. A schematic diagram describing the experimental process for synthesizing $CoFe_2O_4/(MnFe_2O_4)_n$ and $MnFe_2O_4/(CoFe_2O_4)_n$ core/shell IONPs with different shell layers.

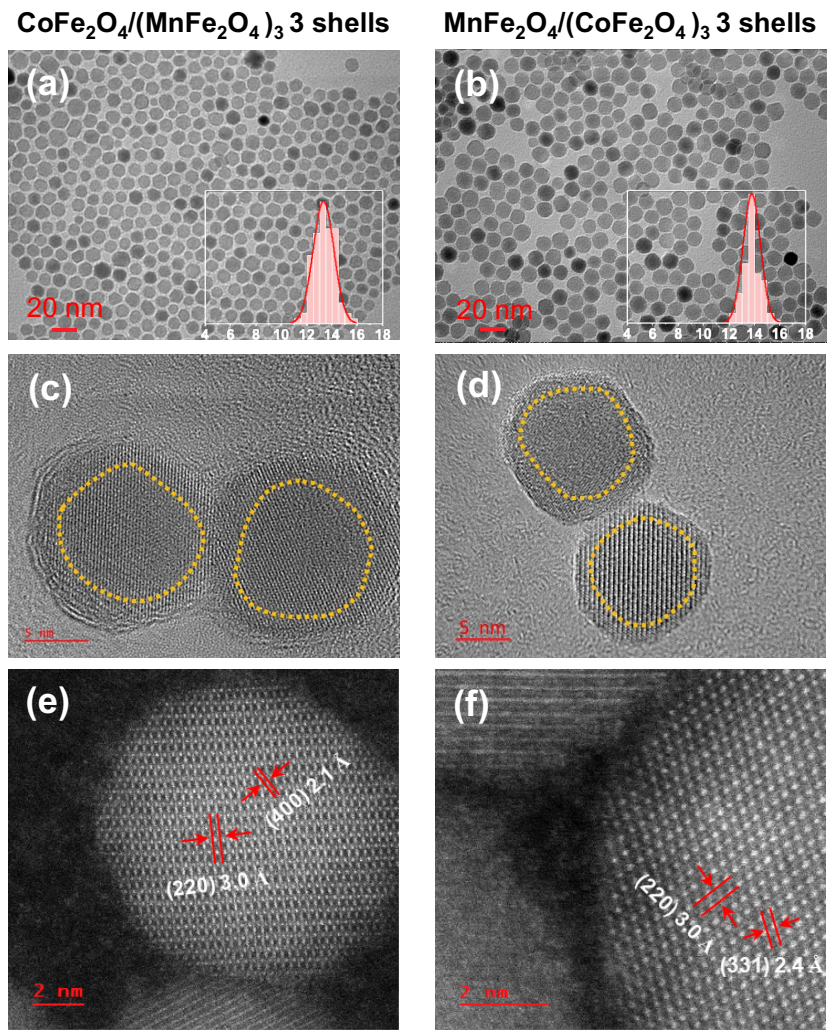

Figure 2. TEM images of (a) CoFe$_2$O$_4$/(MnFe$_2$O$_4$)$_3$ and (b) MnFe$_2$O$_4$/(CoFe$_2$O$_4$)$_3$ core/shell IONPs with 3 shell layers., and the corresponding HRTEM (c and d) and HRSTEM (e and f) of the core/shell IONPs. Insets in (a) and (b) display the particle size histograms of the core/shell IONPs. Yellow lines in (c) and (d) highlight the interface boundaries between CoFe$_2$O$_4$ and MnFe$_2$O$_4$. Red lines in (e) and (f) indicate the atomic lattice fringes of the core/shell IONPs.

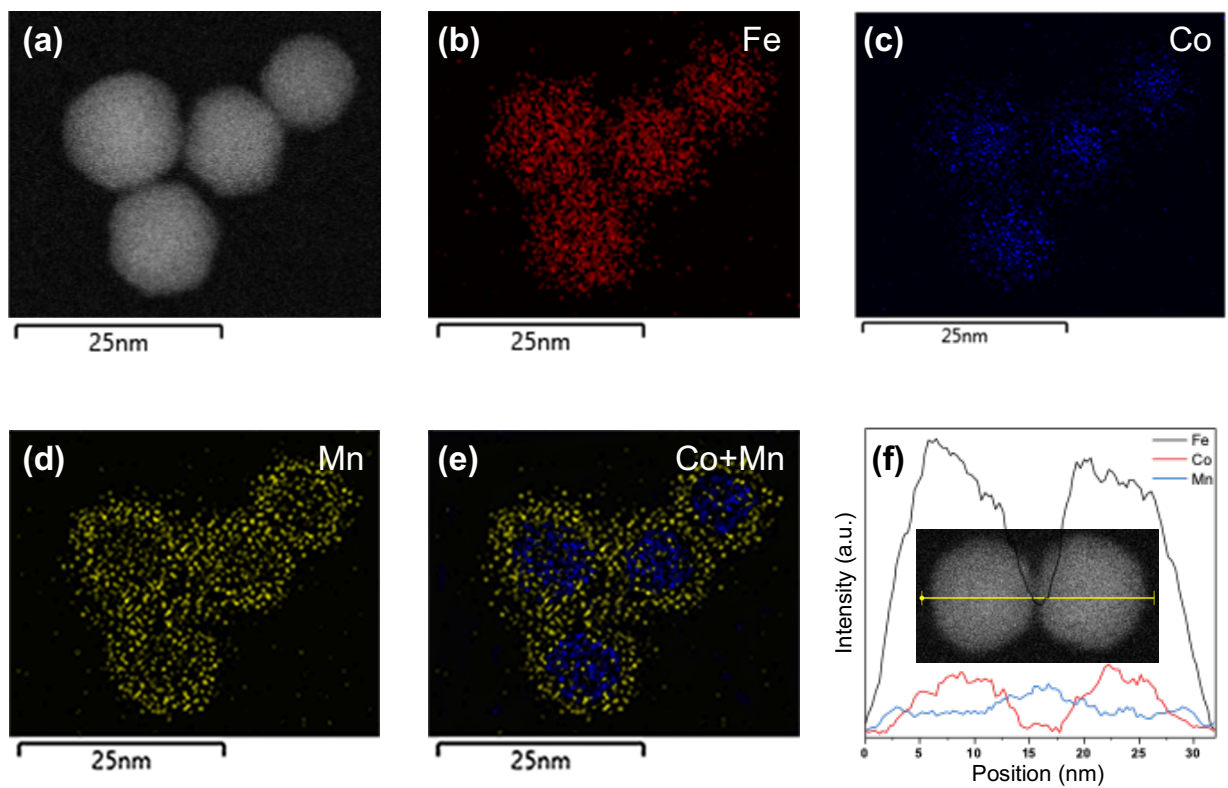

Figure 3. (a) HAADF-STEM images of the synthesized $CoFe_2O_4/(MnFe_2O_4)_3$ core/shell IONPs with 3 shell layers. (b)-(d) Chemical composition mapping of Fe, Co, Mn colored with red, blue and yellow, respectively. (e) Merged image of Co-rich core and Mn-rich shell in the core/shell IONPs. (f) Line-scanned EDS profile of Fe, Co, Mn for two nanoparticles.

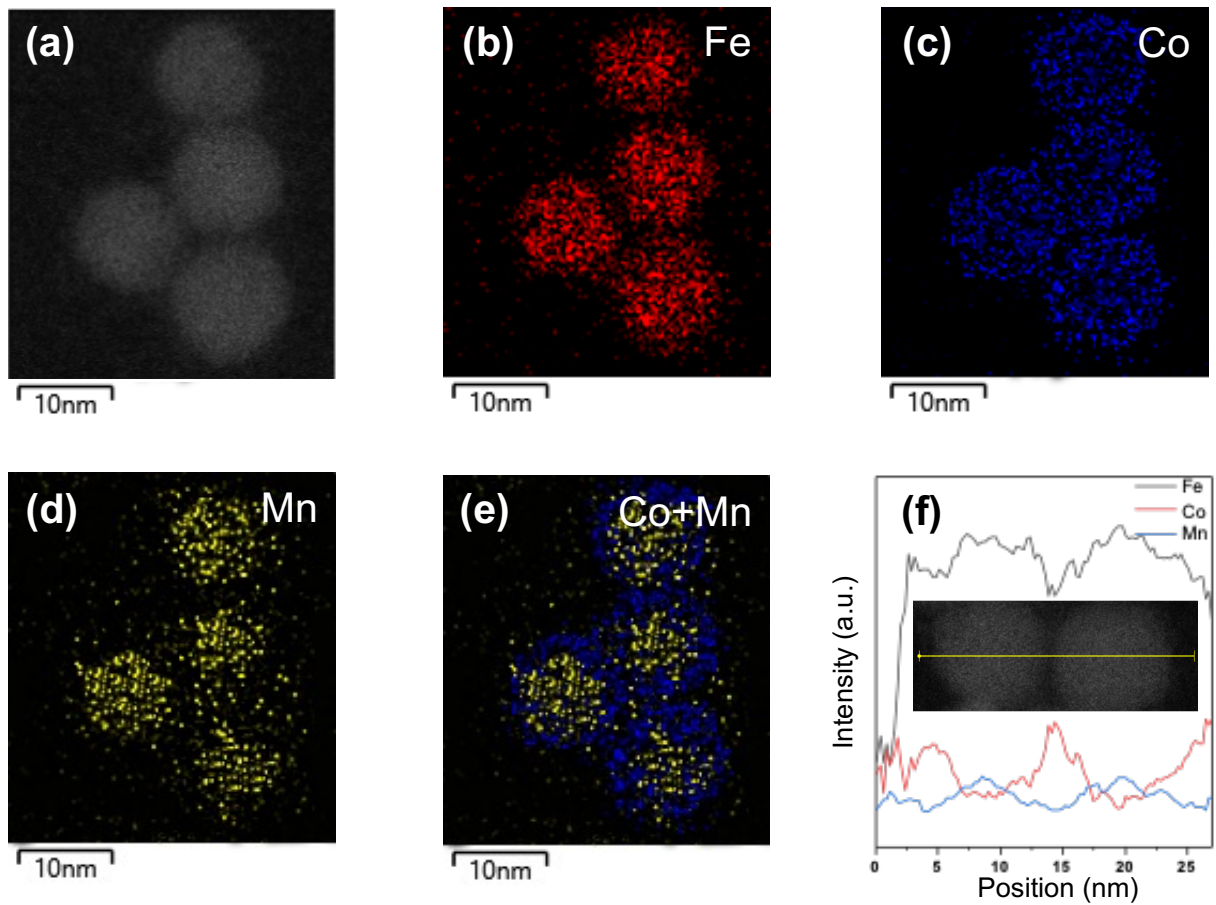

Figure 4. (a) HAADF-STEM images of the synthesized MnFe$_2$O$_4$/(CoFe$_2$O$_4$)$_3$ core/shell IONPs with 3 shell layers. (b)-(d) Chemical composition mapping of Fe, Co, Mn colored with red, blue and yellow, respectively. (e) Merged image of Mn-rich core and Co-rich shell in the core/shell IONPs. (f) Line-scanned EDS profile of Fe, Co, Mn for two nanoparticles.

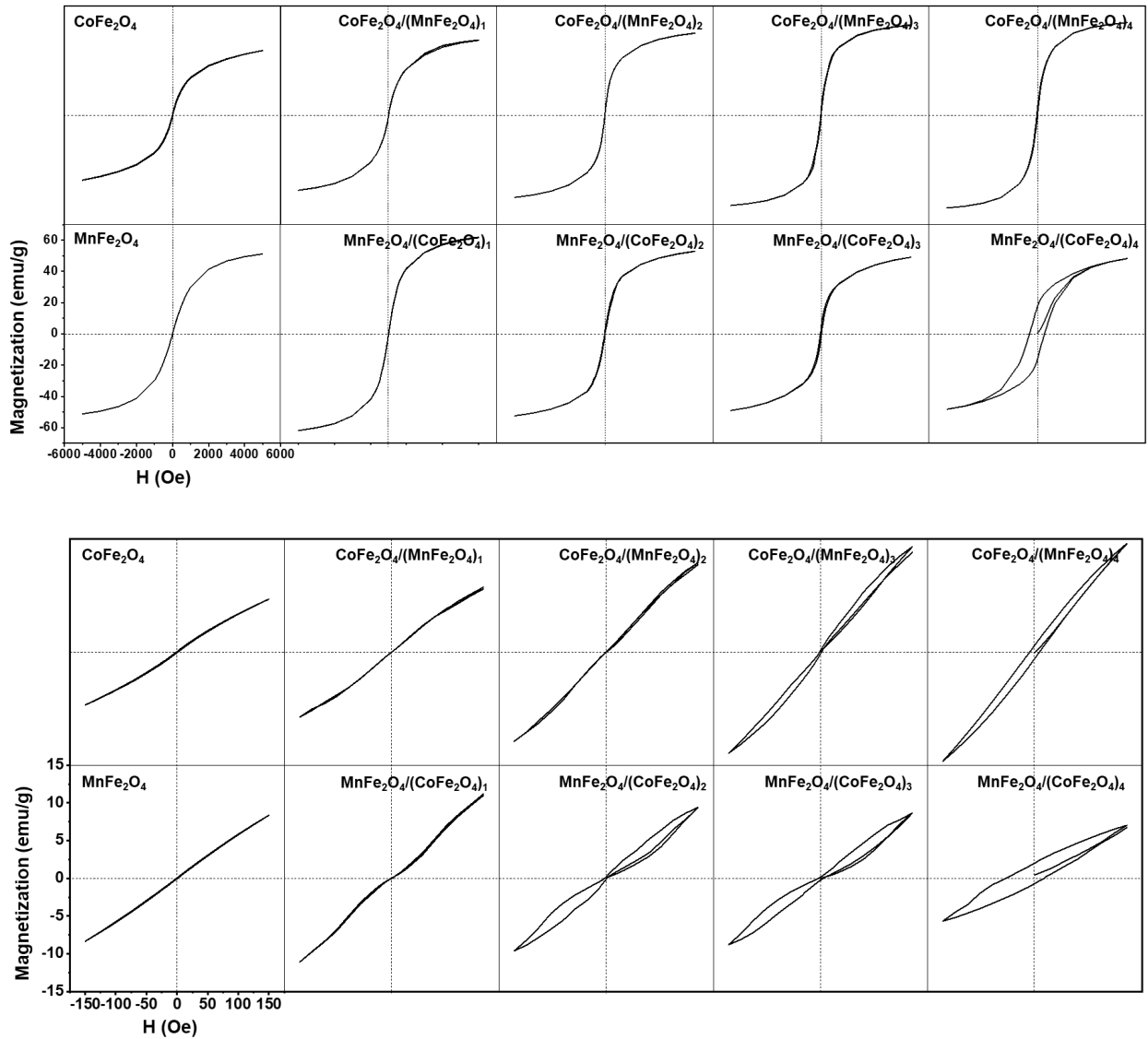

Figure 5. DC major (top), and minor (bottom) M-H loops of single phase $CoFe_2O_4$, $MnFe_2O_4$ IONPs, $CoFe_2O_4/(MnFe_2O_4)_n$ and $MnFe_2O_4/(CoFe_2O_4)_n$ core/shell IONPs measured at the sweeping field of ±5000 Oe, and ±150 Oe, respectively.

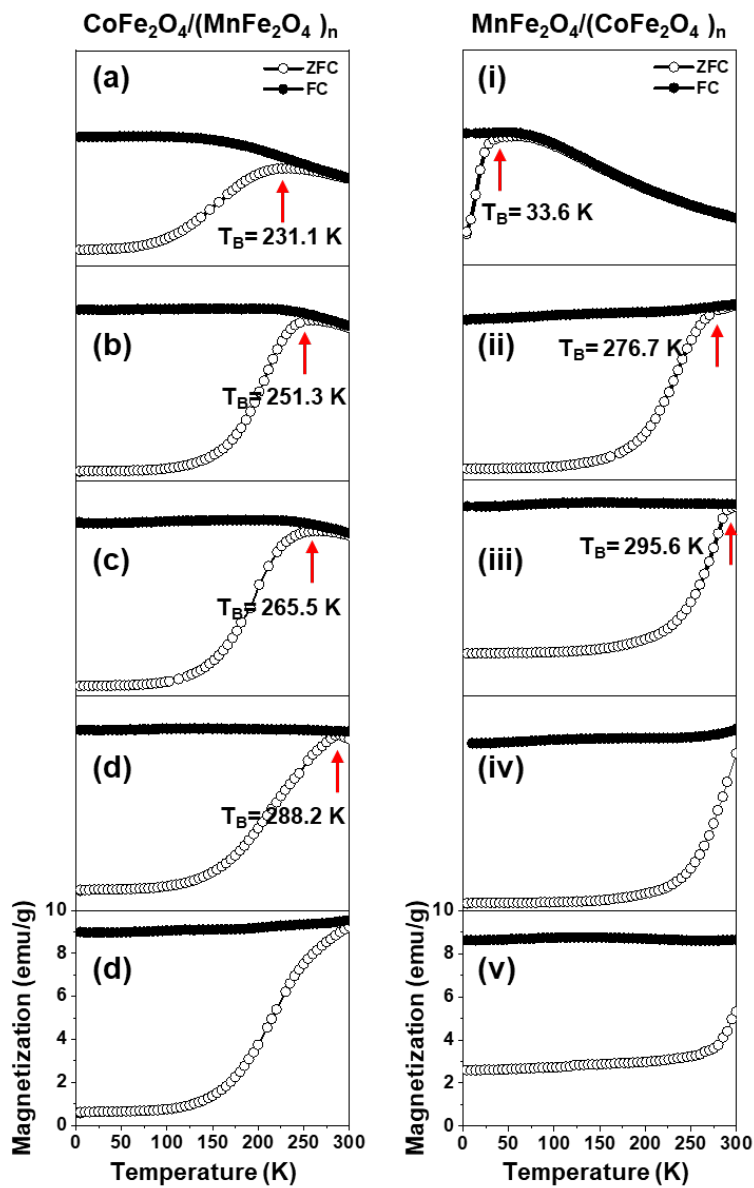

Figure 6 temperature-dependent ZFC (open circle)-FC (filled circle) curves of single phase (a) $CoFe_2O_4$, (i) $MnFe_2O_4$, (b-d) $CoFe_2O_4/(MnFe_2O_4)_n$ and (ii-v) $MnFe_2O_4/(CoFe_2O_4)_n$ core/shell IONPs with different shell layers. The temperature was changes from 4.2 K to 304 K and a constant excitation field of 150 Oe was applied during the FC measurement.

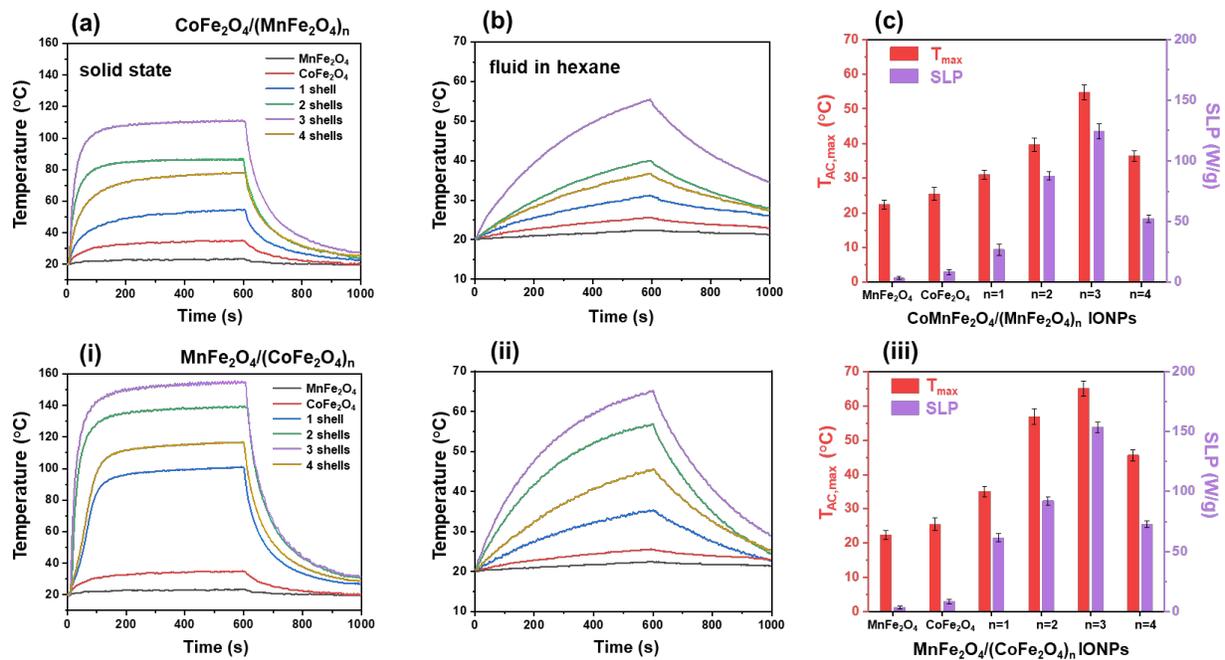

Figure 7. AC heat induction characteristics of $CoFe_2O_4/(MnFe_2O_4)_n$ and $MnFe_2O_4/(CoFe_2O_4)_n$ core/shell IONPs measured at the $H_{appl}$ = 140 Oe and $f_{appl}$ = 100 kHz. (a)-(c) represent AC magnetically-induced heating curves of the solid-state $CoFe_2O_4/(MnFe_2O_4)_n$ core/shell IONPs, core/shell IONP nanofluids (dispersed in hexane), and the corresponding $T_{AC,max}$ and the calculated SLP values, respectively. (i)-(iii) represent AC magnetically-induced heating curves of the solid-state $MnFe_2O_4/(CoFe_2O_4)_n$ core/shell IONPs, core/shell IONP nanofluids (dispersed in hexane), and the corresponding $T_{AC,max}$ and the calculated SLP values, respectively.

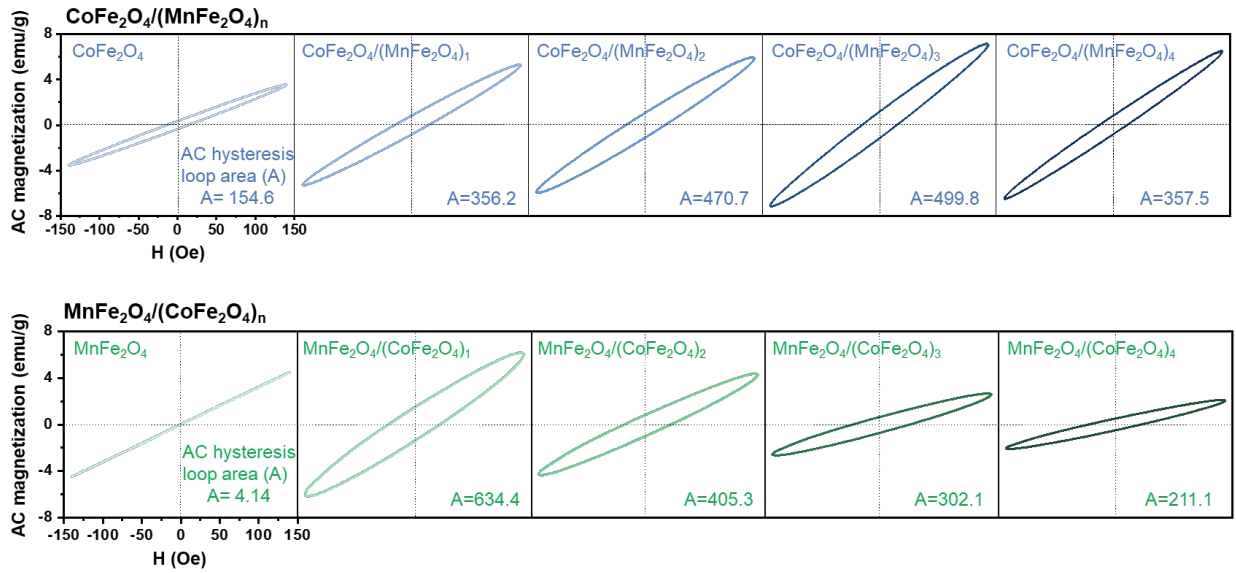

Figure 8 AC hysteresis loops of single phase $CoFe_2O_4$, $MnFe_2O_4$, $CoFe_2O_4/(MnFe_2O_4)_n$ and $MnFe_2O_4/(CoFe_2O_4)_n$ core/shell IONPs measured at the $H_{appl}$ = 140 Oe and $f_{appl}$ = 100 kHz.

Table 1 Physical parameters of $CoFe_2O_4/(MnFe_2O_4)_n$ and $MnFe_2O_4/(CoFe_2O_4)_n$ core/shell IONPs analyzed by DC M-H loops and ZFC-FC measurements. Internal bias field $H_{int}$ was calculated by $H_{int} = \frac{H_{c-}+H_{c+}}{2}$, uniaxial anisotropy constant $K_u$ was determined by the measured Blocking temperature ($T_B = \frac{K_u V}{k_B \cdot \ln(\tau_m/\tau_0)}$), and Neel relaxation time constant $\tau_N$ was numerically calculated by $\tau_N = \tau_0 \exp\left(\frac{K_u V}{k_B \cdot T}\right)$.

| IONPs | $H_c$ (Oe) | $H_{int}$ (Oe) | $M_s$ (emu/g) | $T_B$ (K) | $K_u$ ($\times 10^4$ J/m$^3$) | $\tau_N$ ($\times 10^{-8}$ s) |
|---|---|---|---|---|---|---|
| $CoFe_2O_4$ | 0 | 0 | 41.4 | 231.1 | 9.08 | 1.19 |
| $CoFe_2O_4/(MnFe_2O_4)_1$ | 0 | 0 | 46.3 | 251.3 | 2.34 | 1.49 |
| $CoFe_2O_4/(MnFe_2O_4)_2$ | 0 | 0 | 53.9 | 265.5 | 1.62 | 2.90 |
| $CoFe_2O_4/(MnFe_2O_4)_3$ | 2.02 | -0.25 | 56.3 | 288.2 | 1.58 | 4.27 |
| $CoFe_2O_4/(MnFe_2O_4)_4$ | 7.96 | -0.41 | 58.2 | > 300 | - | - |
| $MnFe_2O_4$ | 0 | 0 | 51.2 | 33.6 | 1.43 | 0.14 |
| $MnFe_2O_4/(CoFe_2O_4)_1$ | 0 | 0 | 62.1 | 276.7 | 2.08 | 1.93 |
| $MnFe_2O_4/(CoFe_2O_4)_2$ | 1.68 | -0.45 | 52.9 | 295.6 | 1.71 | 4.61 |
| $MnFe_2O_4/(CoFe_2O_4)_3$ | 3.73 | -0.53 | 49.1 | > 300 | - | - |
| $MnFe_2O_4/(CoFe_2O_4)_4$ | 31.43 | -16.2 | 48.2 | > 300 | - | - |

**Supplementary**

**Observation of double hysteresis in CoFe$_2$O$_4$/MnFe$_2$O$_4$ core/shell nanoparticles and its contribution to AC magnetic heat induction**

Jie Wang,[a] Hyungsub Kim,[a] Ji-wook Kim,[a] HyeongJoo Seo,[b] Satoshi Ota,[c] Chun-Yeol You,[b] Yasushi Takemura,[d] and Seongtae Bae[a*]

[a]Nanobiomagnetics and Bioelectronics Laboratory (NB$^2$L), Department of Electrical Engineering, University of South Carolina, Columbia, SC, 29208, USA

[b]Department of Emerging Materials Science, Daegu Gyeongbuk Institute of Science & Technology, Daegu, 42988, Republic of Korea

[c]Department of Electrical and Electronic Engineering, Shizuoka University, Hamamatsu 432-8561, Japan

[d]Department of Electrical and Computer Engineering, Yokohama National University, Yokohama, 240-8501, Japan

[*]Corresponding author:
Email: bae4@cec.sc.edu

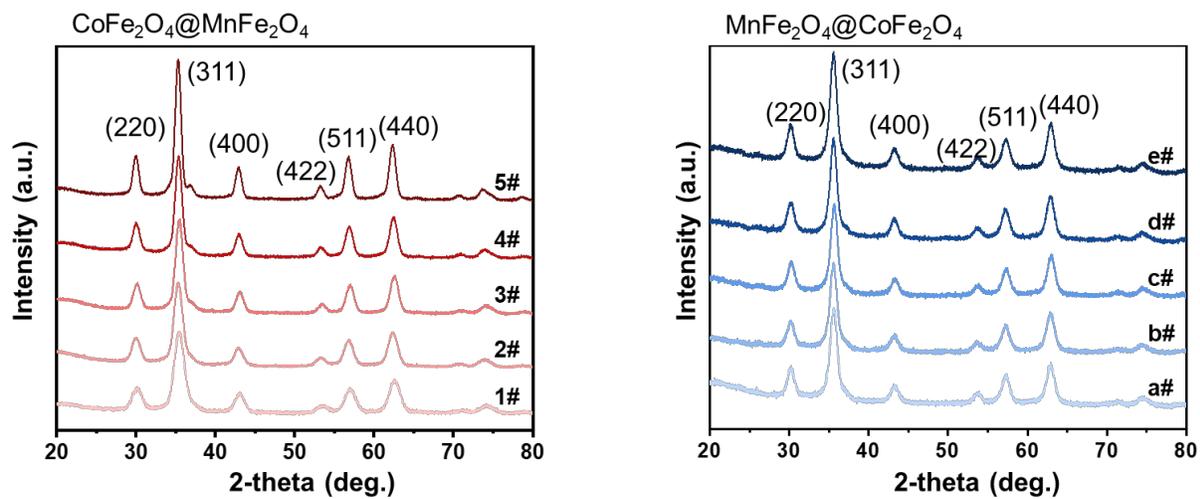

Figure S1. XRD pattern of synthesized core/shell nanoparticles with different shell layers, (a) $CoFe_2O_4/(MnFe_2O_4)_n$ and (b) $MnFe_2O_4/(CoFe_2O_4)_n$.

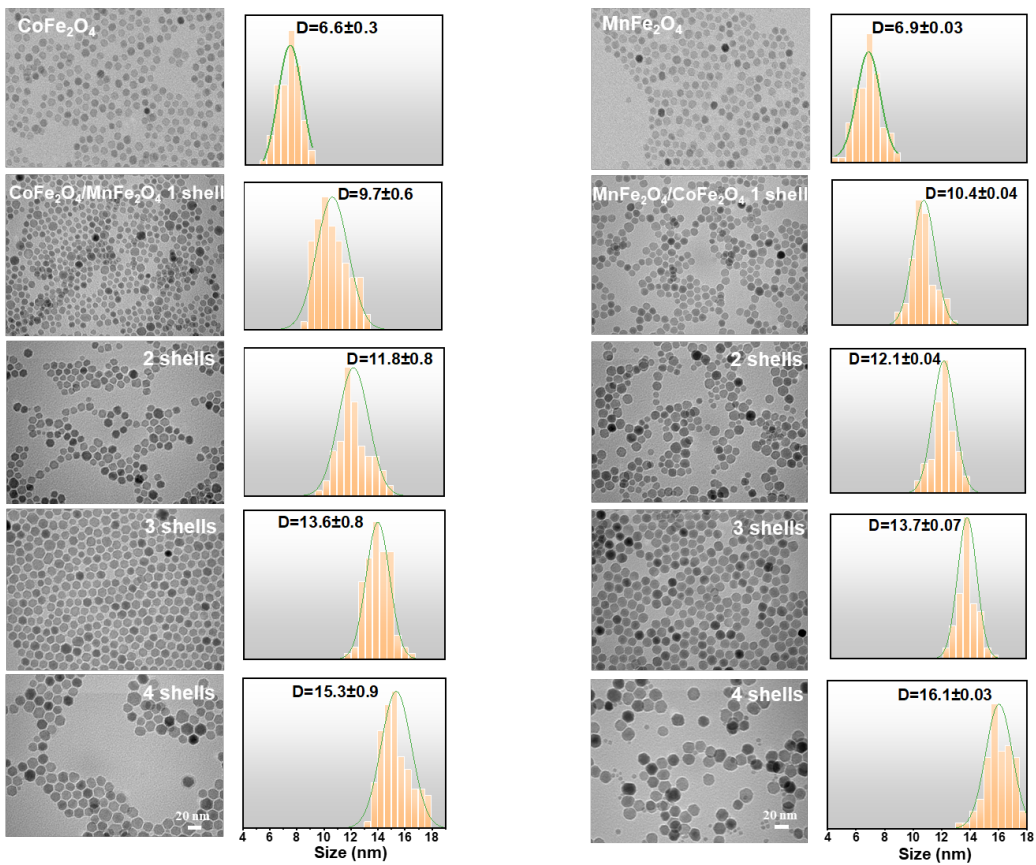

Figure S2. TEM images of $CoFe_2O_4/(MnFe_2O_4)_n$ and $MnFe_2O_4/(CoFe_2O_4)_n$ core/shell IONPs and corresponding the particle size histograms.

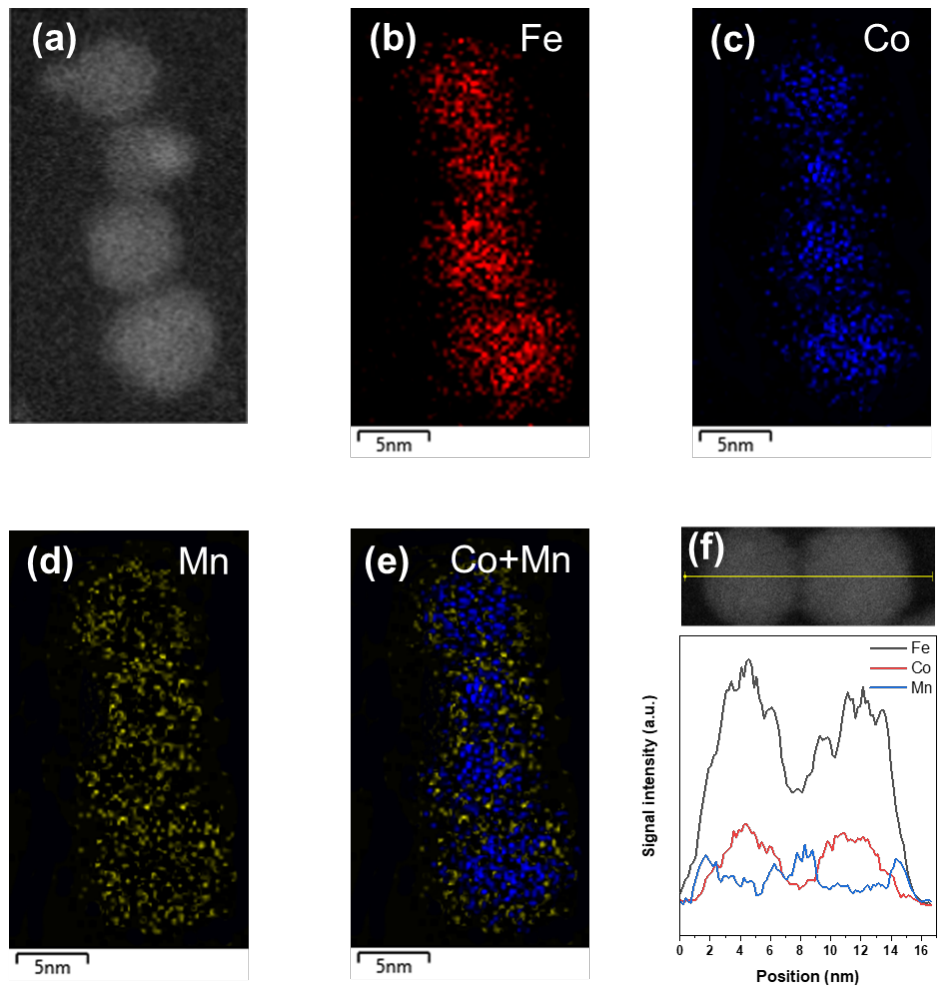

Figure S3. (a) HAADF-STEM images of the synthesized CoFe$_2$O$_4$/MnFe$_2$O$_4$ core/shell IONPs with 1 shell layer. (b)-(d) Chemical composition mapping of Fe, Co, Mn colored with red, blue and yellow, respectively. (e) Merged image of Co-rich core and Mn-rich shell in the core/shell IONPs. (f) Line-scanned EDS profile of Fe, Co, Mn for two nanoparticles.

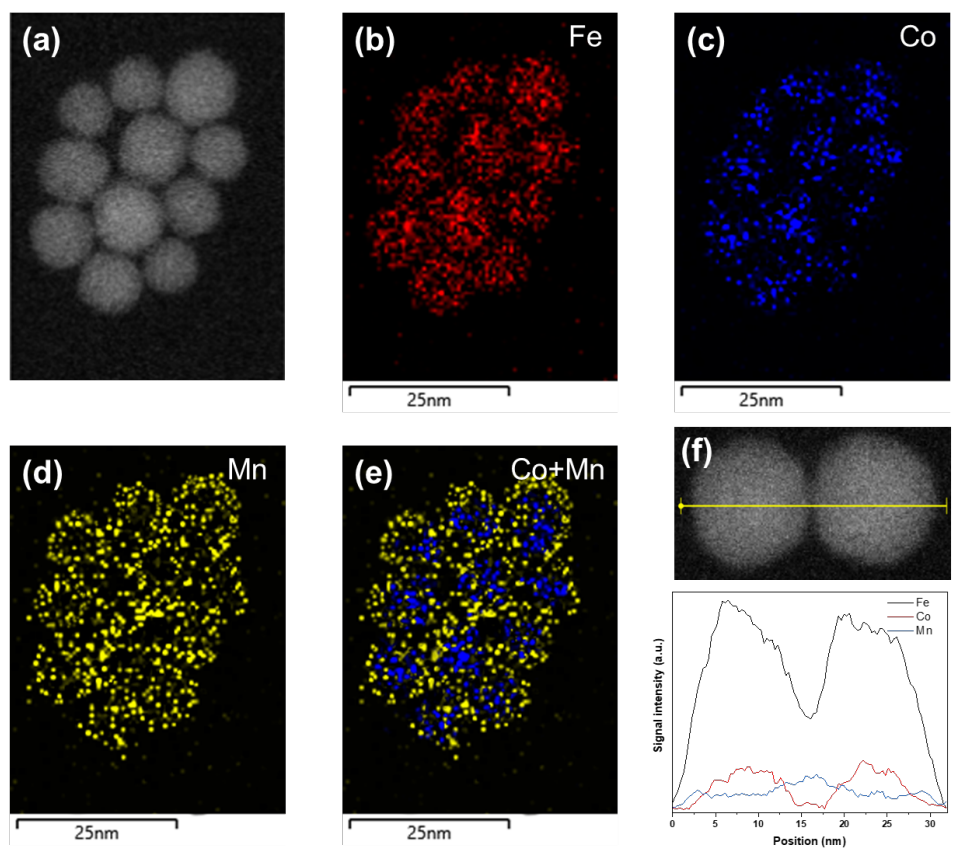

Figure S4. (a) HAADF-STEM images of the synthesized $CoFe_2O_4/(MnFe_2O_4)_2$ core/shell IONPs with 2 shell layers. (b)-(d) Chemical composition mapping of Fe, Co, Mn colored with red, blue and yellow, respectively. (e) Merged image of Co-rich core and Mn-rich shell in the core/shell IONPs. (f) Line-scanned EDS profile of Fe, Co, Mn for two nanoparticles.

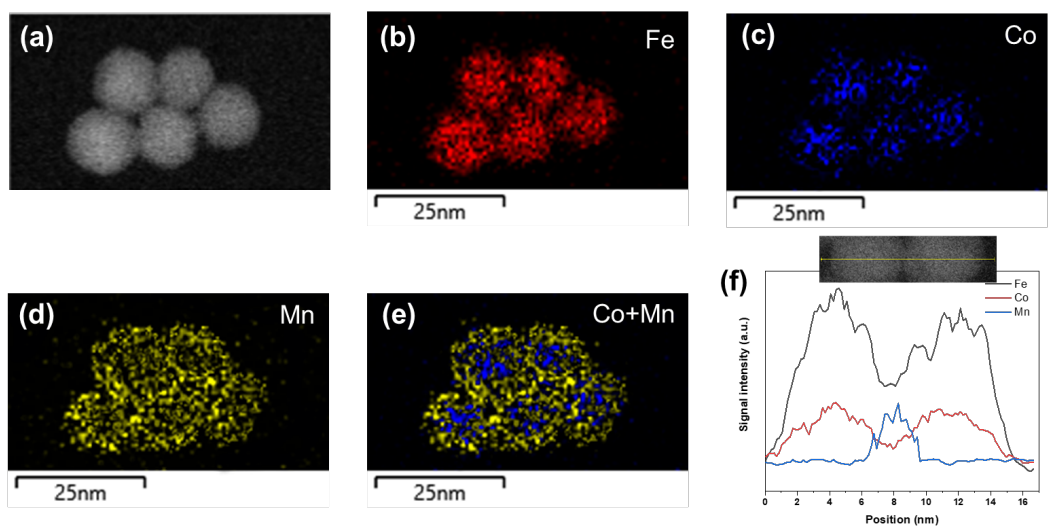

Figure S5. (a) HAADF-STEM images of the synthesized CoFe$_2$O$_4$/(MnFe$_2$O$_4$)$_4$ core/shell IONPs with 4 shell layers. (b)-(d) Chemical composition mapping of Fe, Co, Mn colored with red, blue, and yellow, respectively. (e) Merged image of Co-rich core and Mn-rich shell in the core/shell IONPs. (f) Line-scanned EDS profile of Fe, Co, Mn for two nanoparticles.

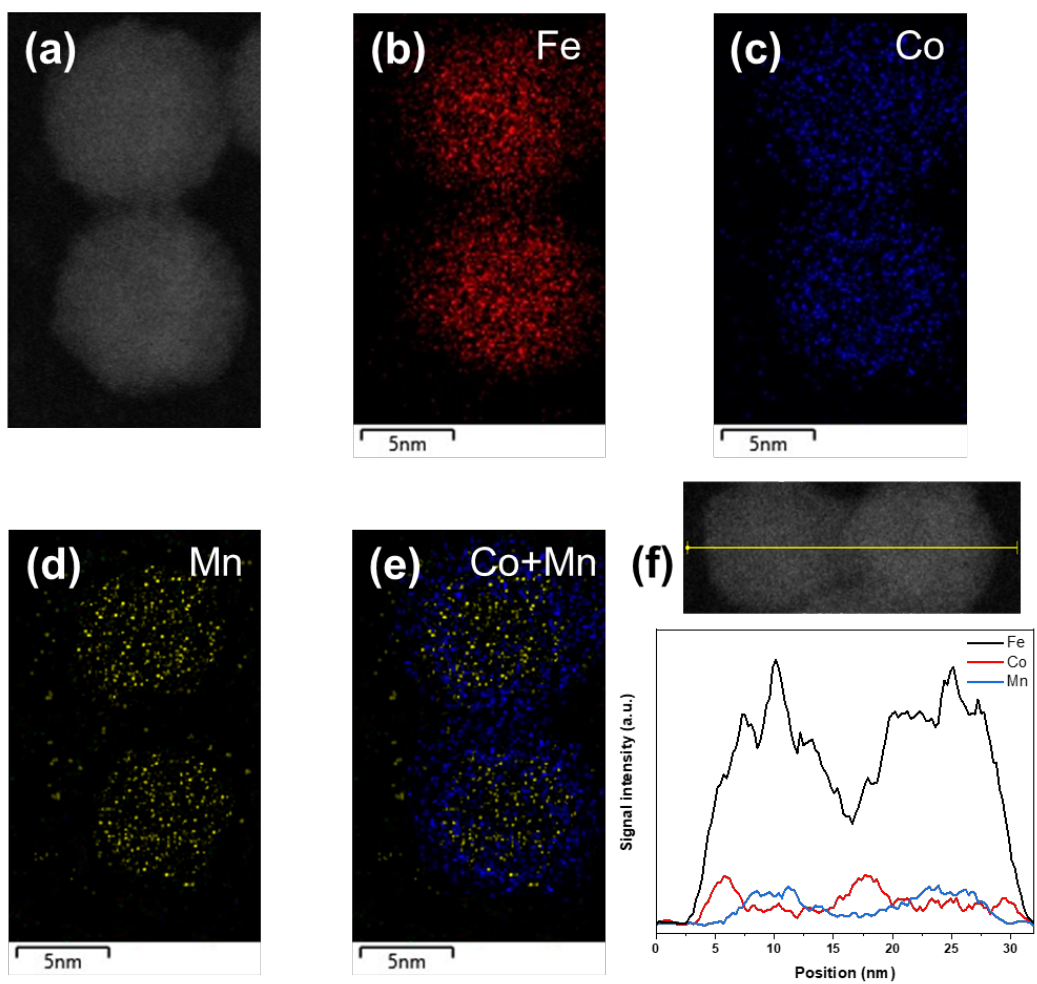

Figure S6. (a) HAADF-STEM images of the synthesized MnFe$_2$O$_4$/CoFe$_2$O$_4$ core/shell IONPs with 1 shell layer. (b)-(d) Chemical composition mapping of Fe, Co, Mn colored with red, blue and yellow, respectively. (e) Merged image of Mn-rich core and Co-rich shell in the core/shell IONPs. (f) Line-scanned EDS profile of Fe, Co, Mn for two nanoparticles.

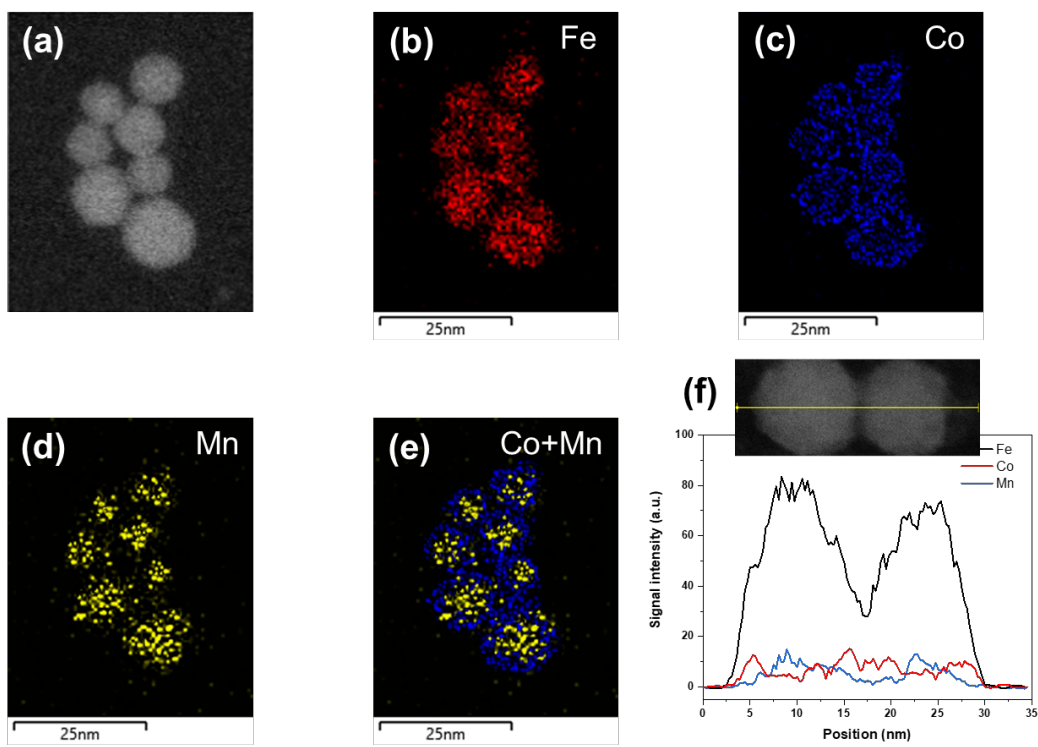

Figure S7. (a) HAADF-STEM images of the synthesized MnFe$_2$O$_4$/(CoFe$_2$O$_4$)$_2$ core/shell IONPs with 2 shell layers. (b)-(d) Chemical composition mapping of Fe, Co, Mn colored with red, blue and yellow, respectively. (e) Merged image of Mn-rich core and Co-rich shell in the core/shell IONPs. (f) Line-scanned EDS profile of Fe, Co, Mn for two nanoparticles.

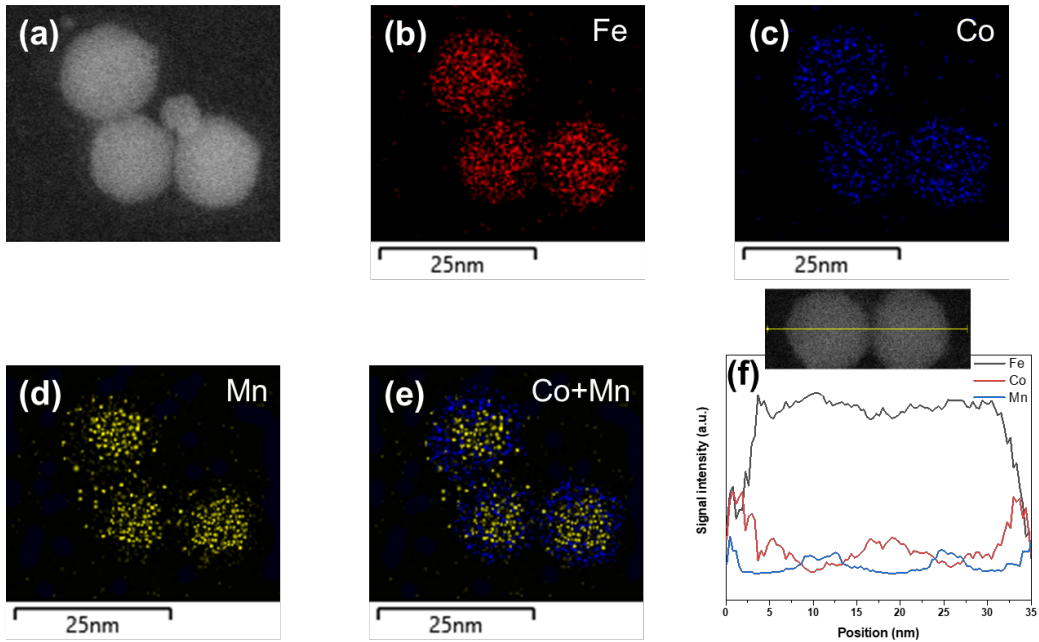

Figure S8. (a) HAADF-STEM images of the synthesized MnFe$_2$O$_4$/(CoFe$_2$O$_4$)$_4$ core/shell IONPs with 4 shell layers. (b)-(d) Chemical composition mapping of Fe, Co, Mn colored with red, blue and yellow, respectively. (e) Merged image of Mn-rich core and Co-rich shell in the core/shell IONPs. (f) Line-scanned EDS profile of Fe, Co, Mn for two nanoparticles.

**Micromagnetic simulation of the magnetic spin state using object oriented micromagnetic framework (OOMMF)**

The micromagnetic simulation of $CoFe_2O_4/MnFe_2O_4$ and $MnFe_2O_4/CoFe_2O_4$ core/shell IONPs was performed with 3D OOMMF. The evaluation of magnetization reversal was obtained by solving Landau-Lifshitz-Gilbert (LLG) equation

$$\frac{d\vec{M}(\vec{r},t)}{dt} = -\gamma \vec{M}(\vec{r},t) \times \vec{H}_{eff} - \frac{\gamma\alpha}{M_s} \vec{M}(\vec{r},t) \times \vec{M}(\vec{r},t) \times \vec{H}_{eff} \qquad (1)$$

where $\vec{M}(\vec{r},t)$ is magnetization vector, $\vec{H}_{eff}$ is effective magnetic field, $M_s$ is saturation magnetization, $\gamma$ and $\alpha$ are gyromagnetic ratio and damping constant, respectively.

The LLG equation describes the movement of the dynamic magnetization under $\vec{M}(\vec{r},t)$ under the action of an externally applied magnetic field ($\vec{H}_{appl}$) as precession movement of $\vec{M}(\vec{r},t)$ around an effective magnetic field ($\vec{H}_{eff}$), defined as

$$\vec{H}_{eff} = -\frac{1}{\mu_0} \frac{\partial E_{eff}}{\partial \vec{M}} \qquad (2)$$

It is assumed that $E_{eff}$ embeds the energy of an effective magnetic field, which is in turn generally expressed by sum of the four energy fields, including magnetostatic energy, anisotropy energy, exchange energy and Zeeman energy, namely $E_{eff} = E_{ms} + E_{ani} + E_{ex} + E_{Zee}$. The equilibrium state of the system minimizes the total energy. The resulting dynamics is that the magnetization vector processes around the $\vec{H}_{eff}$ field, loosing energy due to the damping term, eventually leading to an alignment of $\vec{M}$ with $\vec{H}_{eff}$.

Table S1 Parameters used for OOMMF simulation

| Simulation parameter | | Value |
|---|---|---|
| Exchange stiffness A (J/m) | | $1.33 \times 10^{-11}$ |
| Uniaxial anisotropy $K_u$ (J/m³) | $CoFe_2O_4$ | $2 \times 10^5$ |
| | $MnFe_2O_4$ | $5 \times 10^3$ |
| Saturation magnetization $M_s$ (A/m) | $CoFe_2O_4$ | $4 \times 10^5$ |
| | $MnFe_2O_4$ | $7 \times 10^5$ |
| Initial magnetization axis $m_0$ | $CoFe_2O_4$ | (110) |
| | $MnFe_2O_4$ | (100) |
| Particle size (nm) | Core | 6.5 |
| | Shell | 1 |

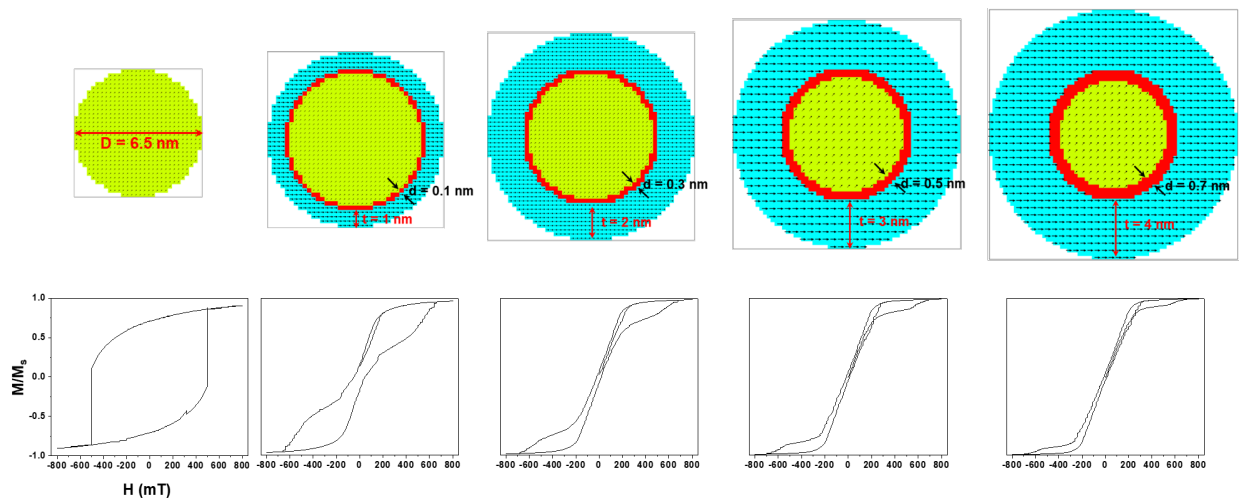

Figure S9 Snapshots of the magnetization configuration at the initial stage for single phase CoFe2O4, and $CoFe_2O_4$/nonmag/$MnFe_2O_4$ core/shell IONPs with different shell thickness (1 ~ 4 nm) and nonmagnetic interlayer thickness (0.1 nm ~ 0.7 nm). The bottom represents the corresponding hysteresis loops after micromagnetic simulation.

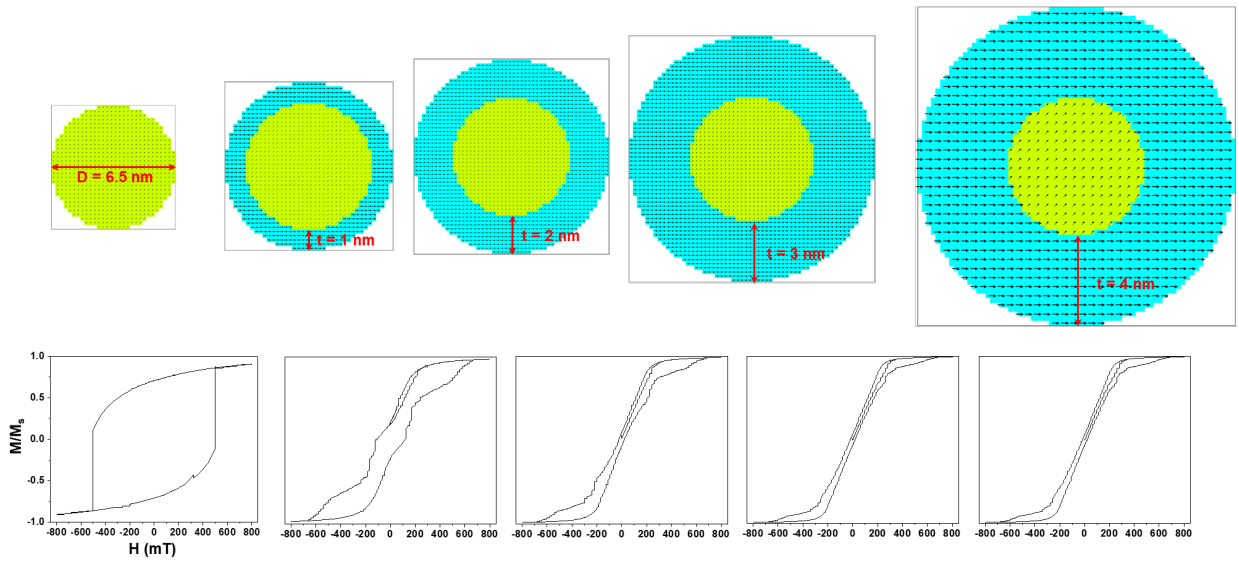

Figure S10 Snapshots of the magnetization configuration at the initial stage for single phase CoFe2O4, and $CoFe_2O_4$/$MnFe_2O_4$ core/shell IONPs with different shell thickness (1 ~ 4 nm). The bottom represents the corresponding hysteresis loops after micromagnetic simulation.

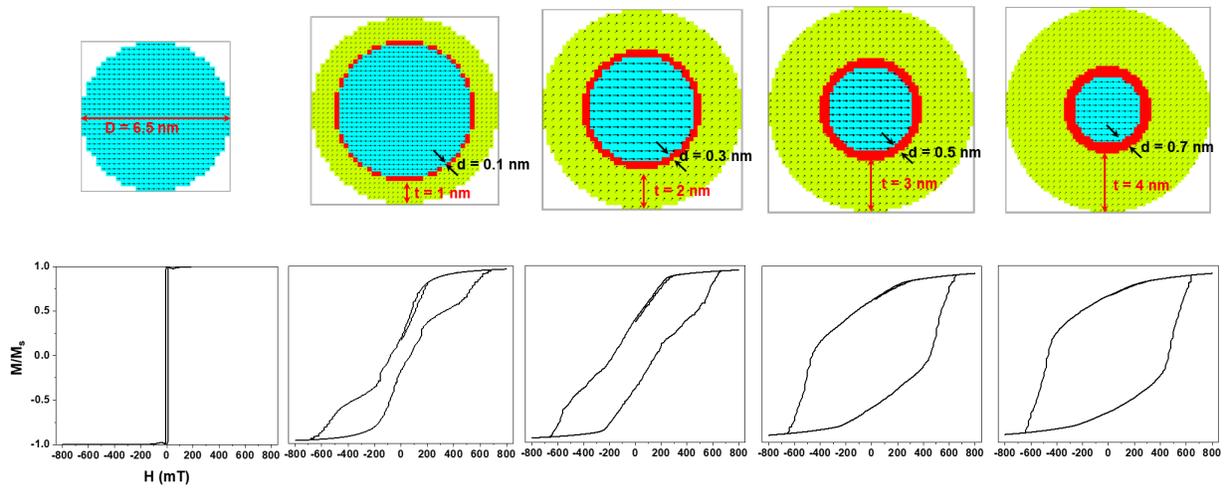

Figure S11 Snapshots of the magnetization configuration at the initial stage for single phase MnFe2O4, and MnFe$_2$O$_4$/nonmag/CoFe$_2$O$_4$ core/shell IONPs with different shell thickness (1 ~ 4 nm) and nonmagnetic interlayer thickness (0.1 nm ~ 0.7 nm). The bottom represents the corresponding hysteresis loops after micromagnetic simulation.

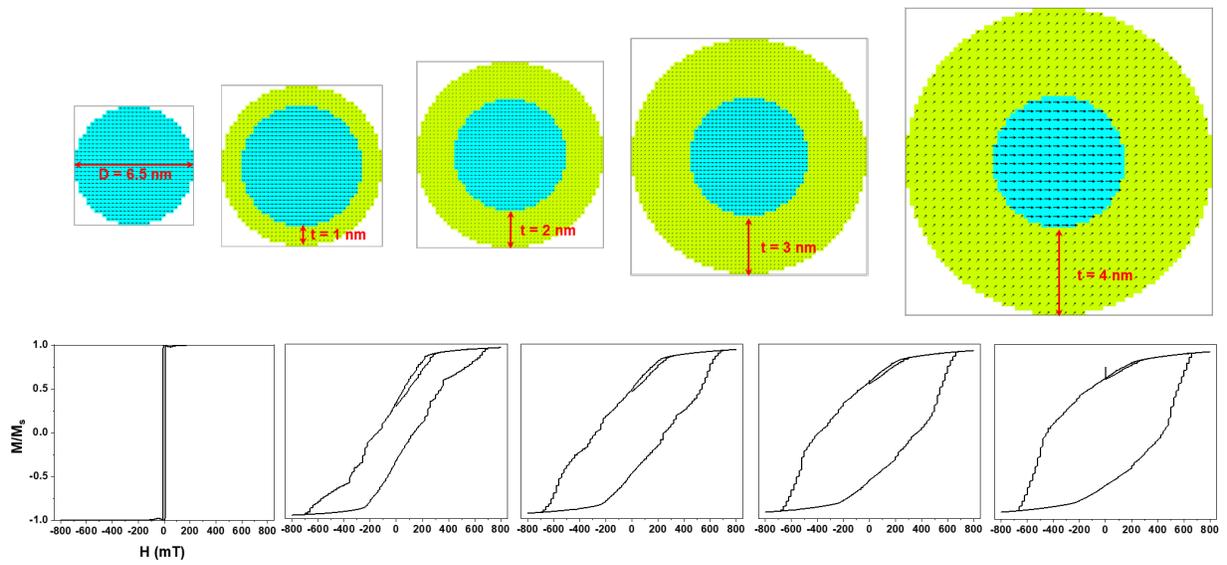

Figure S12 Snapshots of the magnetization configuration at the initial stage for single phase MnFe$_2$O$_4$, and CoFe$_2$O$_4$/MnFe$_2$O$_4$ core/shell IONPs with different shell thickness (1 ~ 4 nm). The bottom represents the corresponding hysteresis loops after micromagnetic simulation.